# A Stochastic Hydrological Model for Regulation of Lake Malawi – Shire River system

Media Sehatzadeh[1],*, Nils Roar Saelthun[2], Jan Atle Roti[1]

[1] Multiconsult ASA, PO Box 265 Skøyen, Oslo, Norway
[2] University of Oslo, PO box 1047 Blindern, Oslo, Norway

**Abstract**

Located in Liwonde in the Southern Region of Malawi, Kamuzu Barrage is the only facility for regulation of water flow down Shire River, which has a major influence on water levels in Lake Malawi. In this article a hydrological model for optimized operation of the Kamuzu Barrage called the *Kamuzu Barrage Operation Model* (KABOM) is presented. This model is a tool for water resource management at the Kamuzu Barrage based on stochastic hydrology. The stochastic model uses water level readings at Lake Malawi and at the Shire River near Liwonde stretching from January 1900 to date. By using historical rating curves at Liwonde, and developing an updated post-2003 flood rating curve, the model can calculate flow at Liwonde for the entire period. The model simulates available water budgets for a range of probabilities based on the Lake Level and flow record over the past 115 years, plus available data on local inflows to the Shire River from its tributaries. The results provide a basis for choosing a release strategy in Kamuzu Barrage, which is used in optimizing the release at the barrage on daily basis. Resulting quantiles display a strong seasonal variation, highest Freewater forecasts within February and March in the rainy season, and the lowest months of June and July in dry season. The results for simulation are controlled against actual Freewater values in the period of December 2016 to October 2017, which demonstrates a satisfactory validation.

# Introduction

The Shire River/Lake Malawi catchment covers the southern part of the East African Rift. The catchment stretches in a north-south direction from the headwaters in Tanzania at 9º S to the confluence with the Zambezi River in Mozambique, at 17º40' S, a total length of 980 km. The most prominent hydrological feature of the catchment is Lake Malawi, the third largest lake in Africa, with a surface area of 28,750 km$^2$. The catchment area of the lake, to the outlet at Mangochi, is 126,500 km$^2$. The climate of the catchment is dominated by the Inter-Tropical Convergence Zone (ITCZ) and its movement north to south, and some influence from the Atlantic by systems moving south through the Congo Basin.[1] These large weather systems are influenced by oceanic large timescale variations like ENSO (El Niño Southern Oscillation), which lead to strong variations in seasonal water availability. The area of Lake Malawi is remarkably large compared to the land catchment with a catchment area/lake area ratio[2] of approximately 3.4 to 1, causing the lake to act as a powerful flow attenuator through its large storage capacity. The water balance and the water level in the lake are thus very sensitive to variations and trends in precipitation and evaporation. Rainfall is highly seasonal[3], the dry season being from May to November. The length of the wet season varies depending on location, ending in March in the south and in April in the north of the country. The rainy season starts in the south in November with convective activity over high grounds, followed by more widespread rainfall in December–March. The rainy season tails off towards May, with south-east trade winds. Tropical cyclones occasionally pass over the area if a blocking pattern becomes established in the Mozambique Channel. For example, at Zomba a tropical cyclone produced 710 mm of rainfall over 36 hours during December 1946. Mean annual rainfall over Malawi is 1100 mm, ranging between 800 mm to over 2000 mm, depending on location. Generally, the precipitation diminishes from east to west, by the increasing distance from the Indian Ocean. The temperature varies over the year with maximum temperatures between 28–34º C in October/November and minimum temperatures between 7–16 ºC in June/July.

Located in Liwonde in the Southern Region of Malawi, Kamuzu Barrage is the only facility for regulation of water flow down Shire River, which has a major influence on water levels in Lake Malawi. The barrage is of strategic importance for the country in terms of regulating water supply to hydropower plants, irrigations schemes and drinking water as well as flood control. The Kamuzu Barrage has been in use since 1965 and is now in need of technical upgrading. This includes increasing the regulation capacity for the water level of Lake Malawi by raising the top of the gates. Kamuzu Barrage is the only existing facility to regulate the flow in Shire River and, to a certain degree, influence, or regulate the water levels of Lake Malawi. The Kamuzu Barrage is one component of various measures to achieve targets within the Lake Level Control management program. The major intended functions are:
- o  Regulate water flow in Shire River to meet demands from down-stream water users;
- o  Regulate water levels upstream to meet environmental and socio-economic requirements;
- o  Contribute to control the water level in Lake Malawi;
- o  Contribute to control floods.

Literature review reveals a few number of studies on the water balance of Lake Malawi/Shire River system available to this date. Many of the existing literatures dates back to 1960s through 1980s. A notable publication is the historical summary published by Drayton[4] in 1984 which focuses on the analysis of the water balance of this system and summarizes much of the early works prior to this publication. In 1998, Nicholson[5] created a comprehensive visualized summary of Lake Malawi's historical water levels dating back to 1800s. Since 2000s there has been a number of articles available in the literature focusing on flow in Shire River [6-9], water balance for Lake Malawi [2, 7, 10, 11] and climate factors affecting the water budget [1, 12, 13].

Funded by the World Bank, through Malawi's Ministry of Agriculture, Irrigation and Water Development, "detailed design of upgraded Kamuzu Barrage" Project was started in 2010 by Multiconsult, and we have also acted as construction supervisors for the project since 2014. As part of this project we developed a hydrological model for optimized operation of the Kamuzu Barrage called the *Kamuzu Barrage Operation Model* (KABOM), which will be thoroughly discussed in this paper. This model is a tool for water resource management at the Kamuzu Barrage based on stochastic hydrology. The stochastic model uses water level readings at Lake Malawi and at the Shire River near Liwonde stretching from January 1900 to date. By using historical rating curves at Liwonde, and developing an updated post-2003 flood rating curve, the model can calculate flow at Liwonde for the entire period. The model simulates available water budgets for a range of probabilities based on the Lake Level and flow record over the past 115 years, plus available data on local inflows to the Shire River from its tributaries. The objective of the developed model is to use stochastic hydrological modelling for:
- o Forecasting water availability at the Kamuzu Barrage for the next twelve months;
- o Forecasting the lake level (Lake Malawi);
- o Forecasting flows downstream of Liwonde; and
- o Forecasting flows through three hydropower plant: Nkula, Tedzani and Kapichira.

The results provide a basis for choosing the release strategy in Kamuzu Barrage, which is used in optimizing release at the barrage every day.

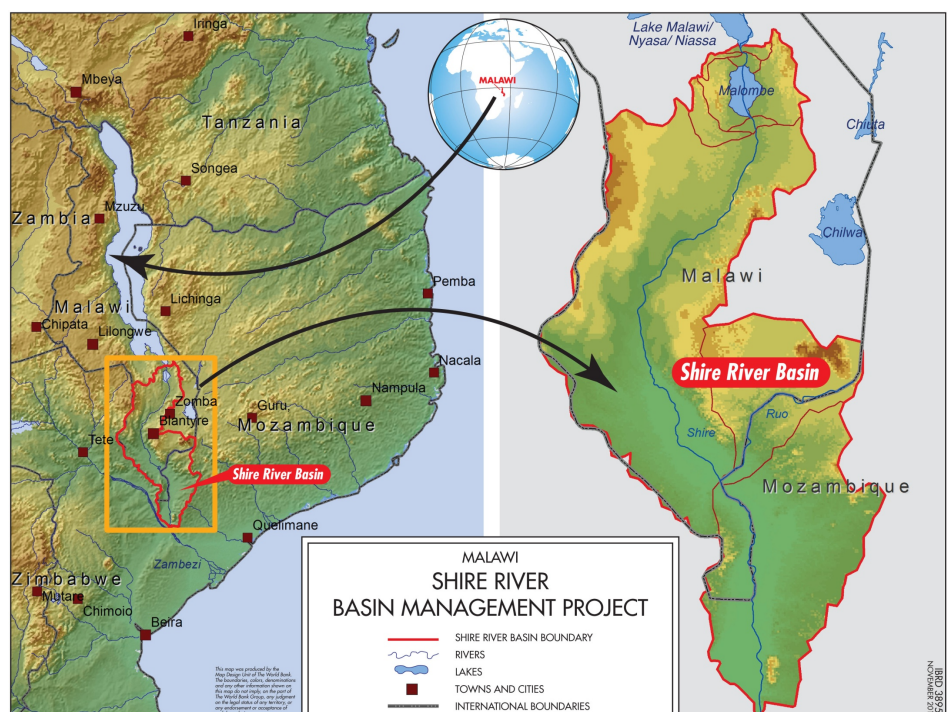

***Figure1:*** *Lake Malawi, Shire River and location of Kamuzu Barrage. The area within the red lines is the Shire river Basin downstream of Lake Malawi. Courtesy of Shire River Basin Management Program project coordinator.*

## Data

Figure 2 shows the water level fluctuations in Lake Malawi since 1800 as appeared in Nicholson[5] (1998). The water level fluctuations have been reliably recorded only since 1896. The values prior to this year are based on Lake Malawi notes of observations from missionaries. The data between years 1800 to 1845 is only indicative corresponding to prolonged low lake levels. The water level values reported in this figure are produced using Shire Valley Project Datum, which is 1.34 m below the National Datum (m.a.s.l.) which is used in this study.

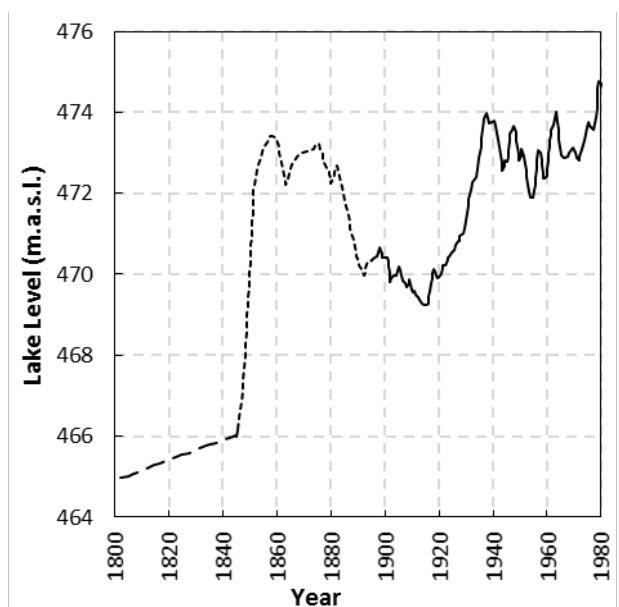

*Figure 2: Estimated water levels of Lake Malawi within the last 200 years. Dashed and dotted data between years 1800 to 1886 is based on observation notes from missionaries. Data between years 1800 to 1845 corresponds to prolonged low lake levels. The water level is produced using Shire Valley Project Datum, which is 1.34 m below the National Datum (m.a.s.l.).*

Figure 3 summarises the Lake Level fluctuations as well as flow data in 1B1 Liwonde in the period of 1900–2016. In most hydrological analyses, hydrometric station 1B1 Liwonde at Kamuzu Barrage (see Figure 4) has represented outflow from Lake Malawi. Actual observations at Liwonde, *i.e.* water level and corresponding flow based on a developed rating curve, stretch back to only 1949. Therefore the unregulated flow at Liwonde in the period of 1900–1948 was estimated by Norconsult[14] (2001) as follows:

$$Q_{Liw} = 69.572 \times (LL - LL_0)^{1.4309} \qquad (1)$$

Where $Q_{Liw}$ is the flow at Liwonde in m³/s, $LL$ is the Lake Level, and $LL_0$ is the sill level. The sill level is assumed at 471.00 m.a.s.l. for the 1900–1908 period, linearly increasing (due to sedimentation) from 471.00 to 474.00 m.a.s.l. in the dry period of 1908–1934, linearly decreasing from 474.00 to 471.50 m.a.s.l. with the increase in outflow in 1935–1948, and stabilized at 471.50 m.a.s.l. afterwards.

From 1949 flow data from 1B1 Liwonde has been used. Due to a large flood which occurred in 2003, the river's cross section is thought to have been changed which subsequently has caused changes in the rating curve that defines the relation between readings of water levels at station 1B1 and the flow. As a part of this study, we carried out control measurements of actual flow in cooperation with the Ministry of Agriculture, Irrigation and Water Development in June 2013. Although not presented in this paper, the results of the control measurements were used for new rating curve at Liwonde, based on which the discharges after 2003 were recalculated.

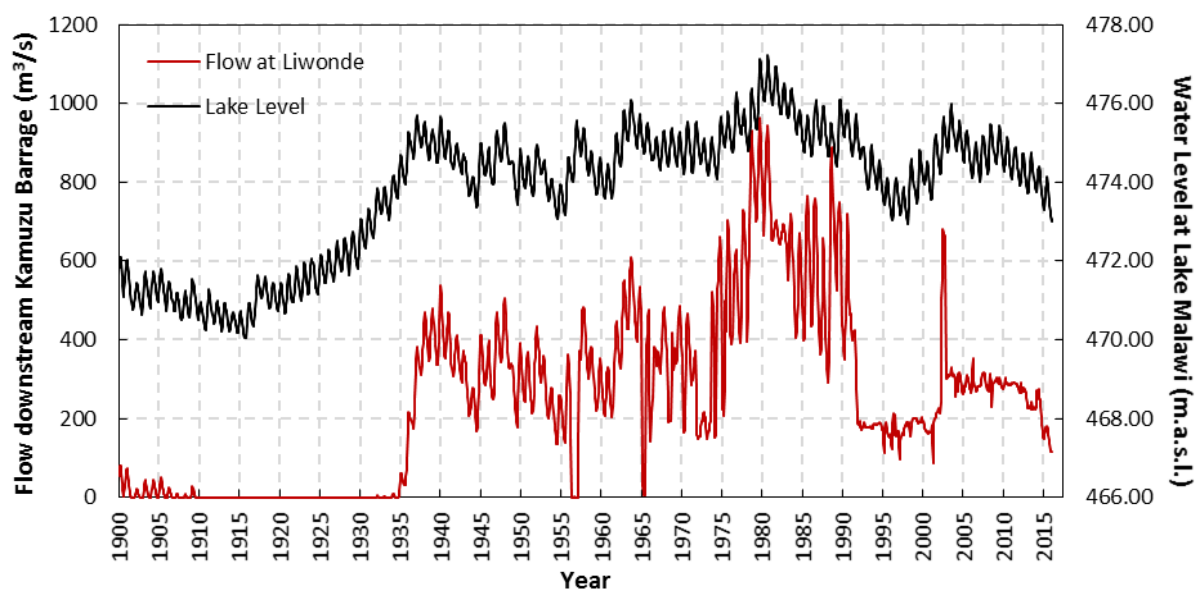

*Figure 3: Monthly record of water level in Lake Malawi (m.a.s.l.) and flow in Shire River at Liwonde ($m^3/s$)*

The Lake's outflow varies with the variations in Lake Level, normally culminating in April. The highest outflow on historical record was in April 1980, approximately 1000 $m^3/s$, with a corresponding Lake Level of 477.25 m.a.s.l.. As the evaporation from the Lake is independent of the inflow, net yield from the lake catchment can be negative. If that continues over a prolonged period, as in the first part of the last century, the Lake Level can be brought below the outlet sill, and there will be no outflow from the Lake. Negative yield brought the Lake Level down to 469.94 m.a.s.l. on 1915, well below the sill, and the Lake was without outflow until 1935.

In an exceptionally dry period between 1900 and 1908, the lake level dropped below the sill level of the outflow to the Shire River (around 471 m.a.s.l.), ceasing outflow of the Shire River entirely. Until and including 1916 the data is interpolated from maximum and minimum annual levels[14] (Norconsult 2001). The period of little-to-no flow from the Lake into Shire River is documented to have lasted up to 1935 partly caused by a developing beach ridge at the outlet to Shire River. In a dry period in late 1990s, the water level dropped as low as 473 m.a.s.l.. There is a decrease in the trend to a minimum 472.98 m.a.s.l. in 1997 and subsequently an increase to a maximum 475.97 m.a.s.l. in 2003. There is a significant decreasing trend in Lake Level after 2009, dropping to 472.98 m.a.s.l. on December 2016.

Besides from Lake Malawi's outflow, the rest of the catchment forms the Shire local catchment (Shire Basin) (22,530 km²), from the outlet of the Lake (start of the Shire River) to the confluence of the Shire and the Zambezi. The local inflow is characterized by large seasonal variations, peaking in January, and quick and large floods.

The flood regime of the Shire River is complex, as it is determined by two components with very different characteristics: the Lake Malawi floods and local catchments floods. The Lake Malawi floods, as a result of a long-term high yield and consequent build-up of Lake Level, dominate in the upper Shire, including the Kamuzu Barrage. The local flood regime has increasing weight and importance downstream. The Elephant Marsh wetlands in the lower Shire store large amounts of water in a flood situation, and reduce quick local floods. In addition, the Lake Malawi floods are influenced by the operation of the Kamuzu Barrage.

Figure 4 illustrates the tributaries to Shire River with their catchments from their tributaries to Shire River, as well as the gauging stations used in this study, and downstream hydropower plants. Hydrological data used in this study was mainly received from Malawi's Ministry of Agriculture, Irrigation and Water Development. Some data was also retrieved from previous studies performed[14] in 2001 by Norconsult. The four main tributaries to Shire River which are focused at in this study are Lirangwe, Lisungwe, Rivi Rivi, Wamkurumadzi, as demonstrated in Figure 4.

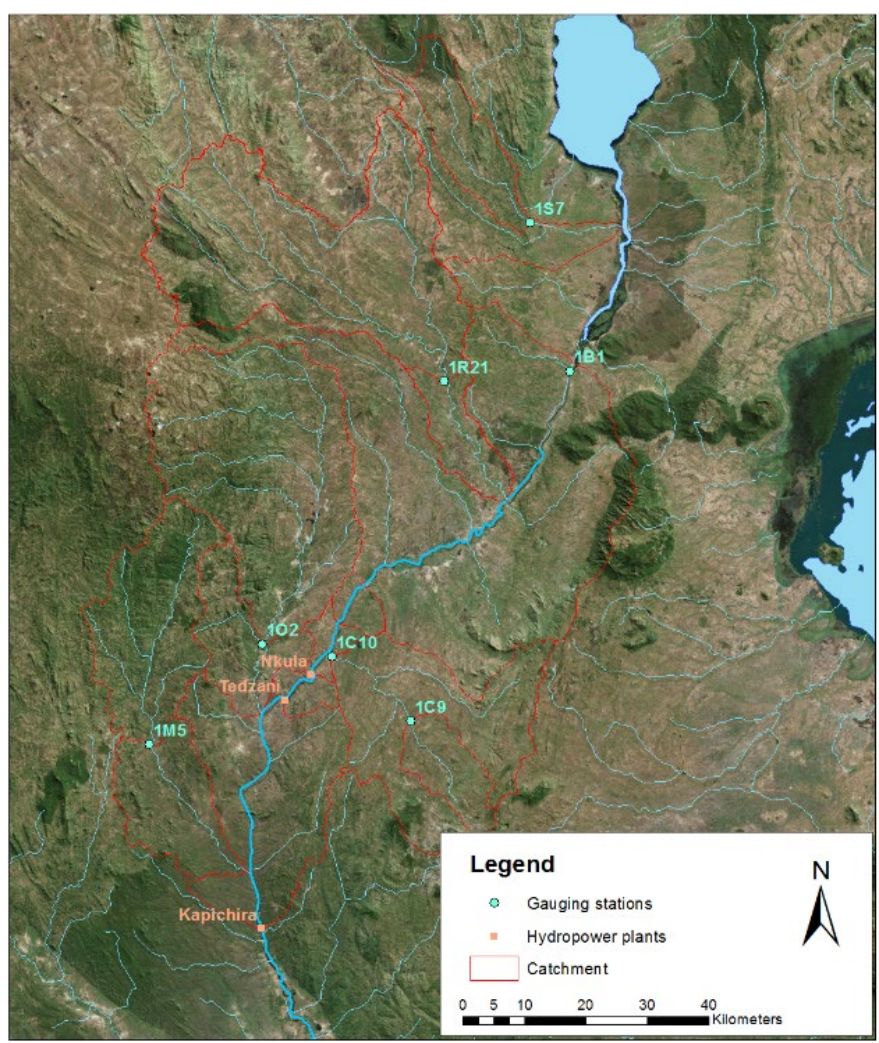

***Figure 4:*** *The tributaries to Shire River with their catchments. The areas surrounded by red lines corresponds to the catchments. The seven cyan notation corresponds to gauging stations from modernization of Shire Basin's hydro-meteorological monitoring system. The three salmon notations corresponds to the existing hydro power plants on the Shire River.*

Mean monthly inflows from the four main tributaries downstream Kamuzu Barrage (Lirangwe, Lisungwe, Rivi Rivi, Wamkurumadzi) are calculated based on a 4–5 year period with available monthly data in the 1980s. The results are summarized in Figure 5. The model combines these historic records with new data implemented from modernization of Shire Basin's hydro-meteorological monitoring system. Figure 4 shows the location of modernized stations, their catchment areas, and the total monthly inflow for their main river based on historical record. As shown in Figure 5, the temporal distribution of flow in the tributaries differs from the temporal distribution of flow in Shire River. The tributaries peak in February–March when the rainfall is most intense.

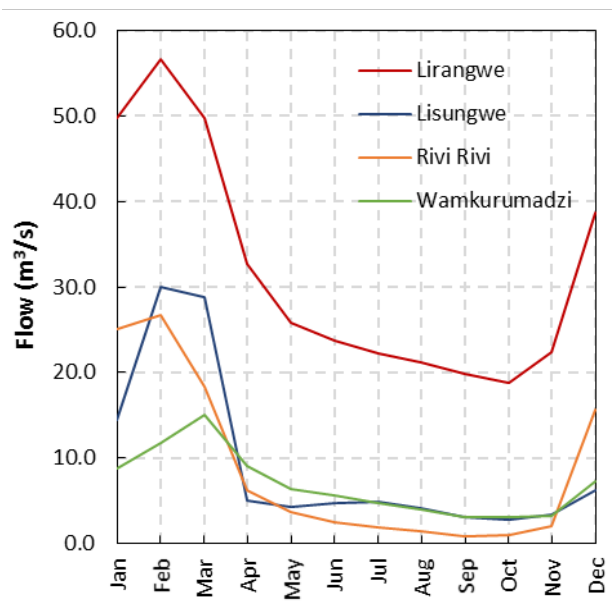

***Figure 5:*** *Monthly flow at the four major tributaries to Shire River based on the historical data.*

## Hydrological Model

The long-term availability of water in the Shire River is strongly dependent on the water budget provided mainly by Lake Malawi. This important variable is termed "Freewater" for historical reasons, *i.e.* the water available for long-term sustainable use out of Lake Malawi. Freewater is defined as:

*Freewater = Land Catchment Runoff + Lake precipitation – Lake evaporation*     (2)

The Freewater values may also be calculated by the following equation:

*Freewater = outflow+ Change in the Lake storage*     (3)

thus:

*Freewater = outflow+ Change in the Lake Level × Lake Area*     (4)

Assuming outflow is equal to flow at 1B1 gauging station at Liwonde (Figures 3 and 4) and given the area of Lake Malawi, the expected Freewater is calculated for each month. In order to use the whole years data, this period is reduced to January 1900 to December 2015, i.e. 1392 months. The results are shown in Figure 6. In any given time series, some pattern of changes may be ascribed to an obvious cause and be explained, leaving a randomly distributed component. Stochastic modeling consists of decomposing a Freewater series into a set of deterministic (*e.g.* general trend, seasonality, autocorrelation) and stochastic (random) terms.

*Freewater = Trend + Periodicity + Autoregressive model + random residual (*$e_t$*)*     (5)

The general trend (linear trend line from Figure 6 data) from 1900 is slightly increasing.

$$Trend_t = 0.229413t + 101.24 \qquad (6)$$

Where t is the months count starting from January 1900.

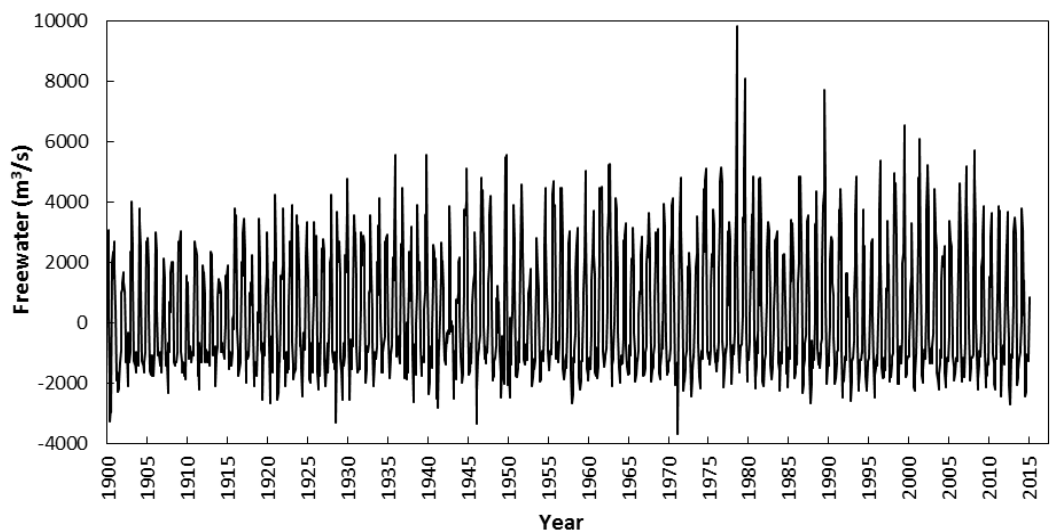

***Figure 6:*** *The monthly Freewater record starting from January 1900.*

*AFRICA 2017 Hydropower Conference, March 2017, Marrakesh, Morocco*

The other deterministic component in Freewater record is the periodicity (also called seasonality or expected seasonal development). Figure 7 shows the average seasonal development calculated based on 115 years data.

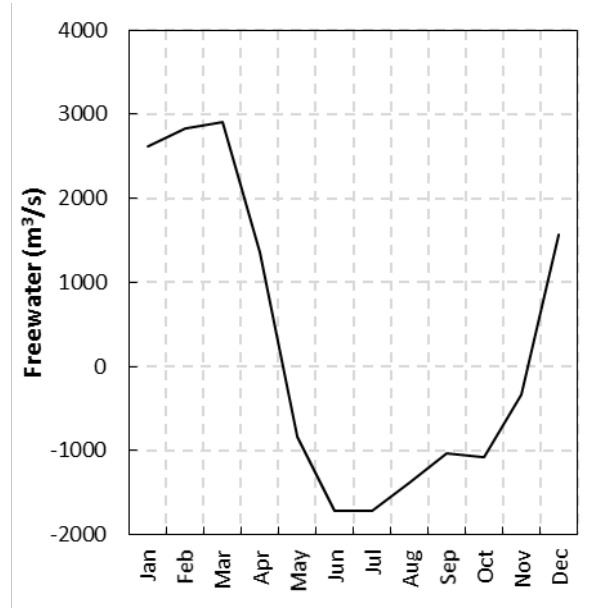

***Figure 7:*** *Average (1900–2015) seasonal development in the Freewater (January–December)*

**Autocorrelation and Stochastic component**

The result of subtracting the trend and seasonality from Freewater is the residual component $E_t$ is shown in Figure 8 and. $E_t$ is autocorrelated which means that the residual Freewater for month *t* is not completely random and depends on the value of $E_{t-1}$.

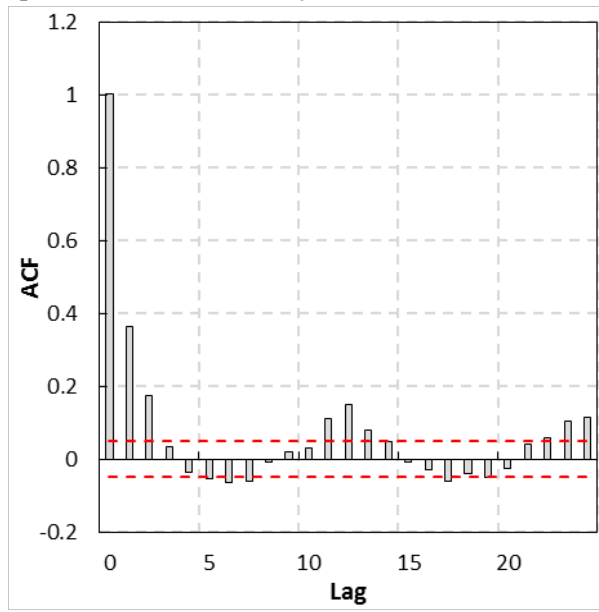

***Figure 8:*** *Autocorrelation in the residual component $E_t$*

The autoregressive model is of second order. This means:

$$AR(2)_t = y_1 \times E_{t-1} + y_2 \times E_{t-2} \qquad (7)$$

In Equation (7), AR and y corresponds to autocorrelation of residual and autocorrelation factor, respectively. Analyzing the series $E_t$ in statistical computing program R, $y_1$ and $y_2$ are estimated respectively as 0.35 and 0.045. $y_2$ will be insignificant and can be ignored when regenerating Freewater.

An autoregressive model is then fitted to the data and later subtracted to obtain independent, identically distributed residuals, $e_t$. These residuals are further described by a normal distribution from which random numbers are drawn. These random numbers exhibit the natural variation of the data series. Therefore, the random residual $e_t = E_t - AR(1)$ is now independent as shown in Figure 9. The histogram of $e_t$ is shown in Figure 10. As this figure shows, $e_t$ has a normal distribution with mean value of –169.5 and standard deviation of 933.6.

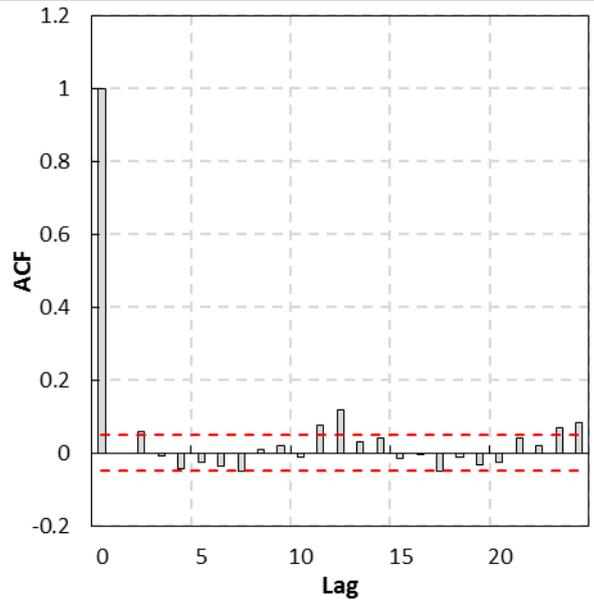

**Figure 9:** *Autocorrelation in random residual $e_t$ ; calculated by the difference between random component ($E_t$) and autoregressive model $AR(1)_t$*

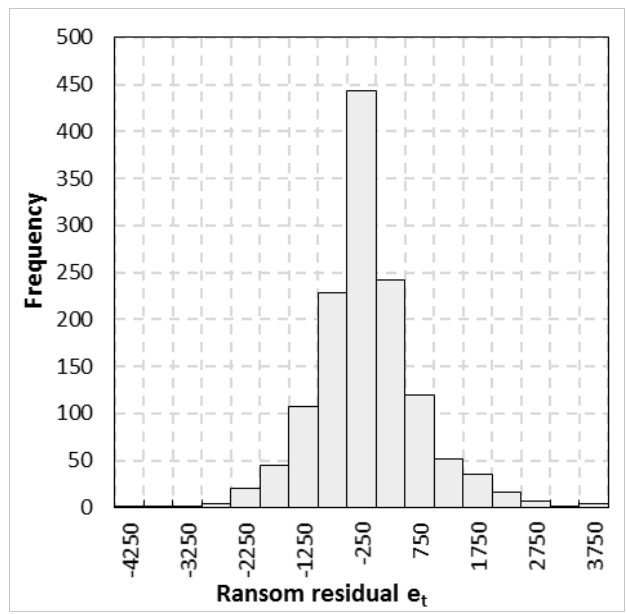

**Figure 10:** *Histogram of random residual $e_t$*

**Operation Model**

In order to create a user-friendly interface, the operation model was developed in Microsoft Excel. The spreadsheet incorporates daily readings from hydrometric stations (principally the water level at Lake Malawi; average of Monkey Bay, Nkhata Bay and Chilumba, *i.e.* Lake Level) and demand data for three downstream hydropower plants (Nkula, Tedzani and Kapichira), irrigation and water supply. Based on these inputs combined with the model's in-built stochastic model, the model produces a recommended release strategy, forecasts in terms of water available at the Barrage (Freewater), water level at Lake Malawi and available flow at the Kamuzu Barrage and the three Hydropower plants downstream. Depending on the inputs and the release strategy, the model will simulate and forecast water levels at Lake Malawi and available flow at the Kamuzu Barrage and the three Hydropower plants downstream. The model's core is the stochastic model and the forecasts produced are a range of Lake Levels with assigned probabilities from 10% to 90%.

**Results and Discussions**

The stochastic series for Freewater is generated by equation (8):

$$Freewater_t = Trend_t + Seasonal\ development_t + Autocorrelation_t + Random\ component_t$$
$$= Average\ for\ 1980–2015\ (311.69\ m^3/s) + Seasonal\ development_t$$
$$+ 0.35 \times (Freewater_{t-1} - Seasonal\ development_{t-1}) + N(\mu = -169.5, \delta = 933.6) \quad (8)$$

An ever-increasing trend over time is unrealistic. Therefore, it is replaced by the average trend in the period 1980–2015. This value is believed to be a realistic replacement even though it is somewhat higher than the 1900–2015 average, which is 261.03 m$^3$/s. For the random component, 100 sets of random numbers were generated, each set having 12 elements. The 100 sets were sorted and shortened to 11 sets for the next 12 months, keeping the nine middle set with the 10, 20, 30, 40, 50 (median), 60, 70, 80 and 90 percentile for each month. Figure 11 presents the resulting Freewater quantiles as monthly time series with labels $q_{10}$, $q_{20}$, etc. The measured values of Freewater for the same period is also presented in Figure 11. Although following the same trend generally, the measured values start from the low Free water quantile of q30, and scan through almost all the quantiles. The highest value of the measured Freewater was in March 2017, corresponding to . exceedance probability of 10% (slightly lower value than those forecasted by q90). The contour illustration of the Freewater prognosis for a 12 months period is presented in Figure 12. The result demonstrate strong seasonal variation in Freewater as a general trend. The most extreme scenarios are the end of the rainy season in the highest quantile (q90) shown by dark red and the end of the dry season in the lowest quantile (q10) shown by dark blue in Figure 12.

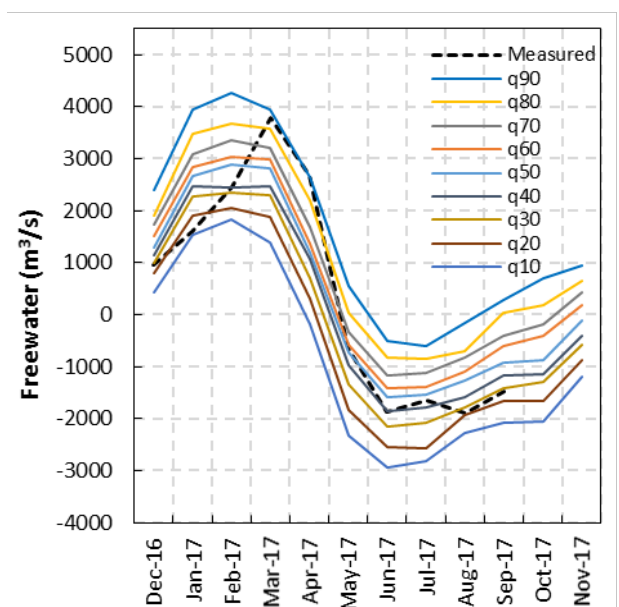

***Figure 11:*** *Stochastic prognosis for Freewater (m$^3$/s) Results for Dec 2016 – Nov 2017*

These large weather systems are influenced by oceanic large timescale variations like ENSO (El Niño Southern Oscillation), and this may be a reason why the Freewater of Lake Malawi display strong variations.

In order to base the calculation at the Kamuzu Barrage, the outflow is equal to flow measured at 1B1 Liwonde (downstream of Kamuzu Barrage). By definition, Freewater is completely independent of regulations of Lake Malawi. However, changes in the catchments (*e.g.* by different land-use) or outtakes for irrigation along Lake Malawi can influence the Freewater available at Kamuzu Barrage.

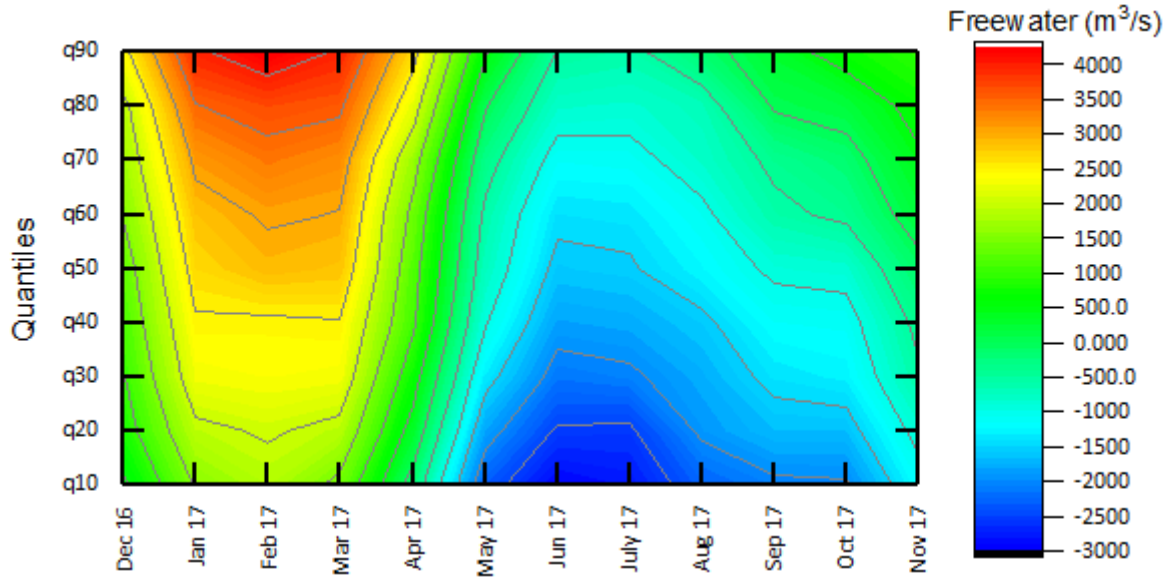

*Figure 12:* The contour visualization of the stochastic prognosis for Freewater ($m^3$/s) (Dec 2016 – Nov 2017)

Given the Freewater and river outflow, the Lake Levels LL(t) is predicted using the following expression if a continuous time series is considered:

$$LL(t) = LL_0 + \int_0^t \frac{[Freewater(t) - Outflow(t)]}{A} \cdot dt \qquad (9)$$

Where LL is the Lake Level at the time *t* and the Freewater and outflow is expressed in $m^3$/s. Simplifying that with time steps of *Δt* equal to 2592000 seconds (for a month) results to this formula for the Lake Level at the time step *t+1*:

$$LL_{t+1} = LL_t + (Freewater_t - Outflow_t) \cdot \Delta t / A \qquad (10)$$

In these calculations Outflow is release at Kamuzu barrage. Lake Level at the first time step is inserted by the user. Figure 13 presents predicted water levels at Lake Malawi, based on an initial Lake Level of 472.99 m.a.s.l. (which is well below the Lowest Regulated Water Level, LRWL) and reduced release at Kamuzu Barrage to constant 110 $m^3$/s.

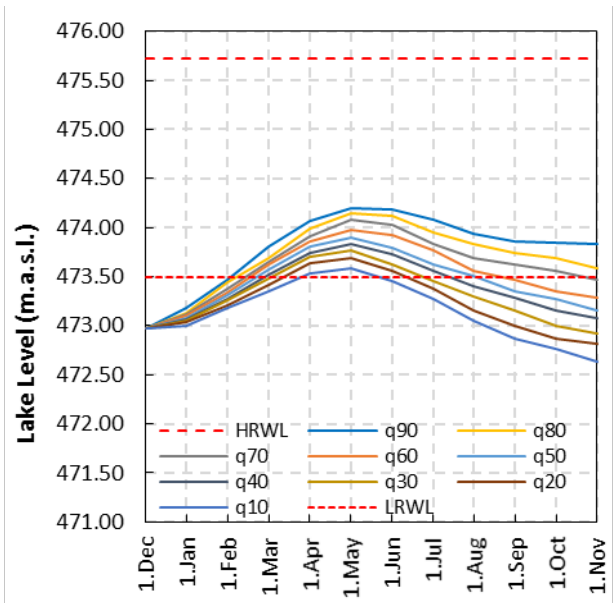

***Figure 13:*** *Stochastic prognosis for Lake Level (m.a.s.l.) at the first day of the month. Results based on Lake Level on December 1$^{st}$ 2016, assuming a constant release of 110 m$^3$/s at Kamuzu Barrage*

At any given point in Shire River downstream of Kamuzu barrage, the flow is the sum of two components:
  i)    The river discharge at Liwonde/Kamuzu Barrage; and
  ii)   The local inflow between Liwonde and the point of interest.

In the natural situation (*i.e.* with all gates of Kamuzu Barrage fully open), the river has a certain natural (unregulated) flow depending on the Lake Level. Here we use a 1-D hydraulic model that we previously developed to simulate the Lake Malawi - Shire River system.[15] Applying this model system results in Figure 14, which is the relation between Lake Level and flow at Liwonde below by a capacity curve.

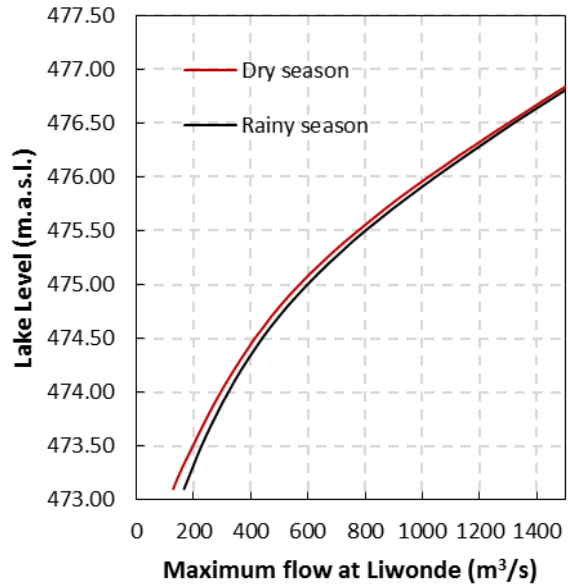

*Figure 14: Capacity curve illustrating the relation between water level at Lake Malawi and flow at station 1B1 (Liwonde) downstream of Kamuzu Barrage for post 2003 period. The difference in capacity in dry season and rainy season is related to damping effect because of storage capacity in Lake Malombe*

Based on the capacity curve, the model produces maximum available flow at the barrage for any given Lake Level. Figure 15 presents predicted available water flow (*i.e.* flow with all barrage gates open) at Kamuzu Barrage, based on starting Lake Level predictions in Figure 13. The results presented in this figure are for the December 2016 – November 2017 period. Note that the results correspond to slightly different Lake Levels from Figure 13, since the excessive release at the barrage will lead to slightly lower Lake Levels (a few centimeters) in the consequent months.

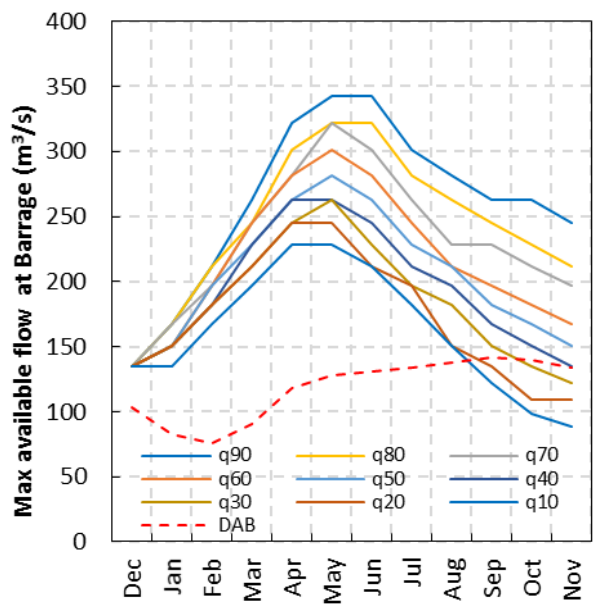

*Figure 15: Stochastic prognosis for available Liwonde flow ($m^3/s$). Results for Dec 2016 – Nov 2017. Recommended release strategy is water saving (Strategy 1), with reduced demands at hydropower plants. The dotted line shows the demand at the barrage and is calculated based on hydropower production of 50% of the installed capacity at power plants, irrigation, drinking water and environmental flow, and local inflows from the tributaries (Figure 5)*

# Conclusion
The model presented in this article has been in use for optimization of water release at Kamuzu Barrage since April 2015. A team of hydrologists in Malawi, continuously use the model for operation of the barrage and management of the water resources in the dry season. The data provided by the operation team is used to update the stochastic model record. Moreover, with the ongoing installation of new hydrological stations on Shire River and its tributaries, the model will receive more accurate, automatic data thus improving accuracy of the model results and the user experience. Optimizing water release at the barrage will be especially important in the upcoming rainy season. Since its first implementation in 2015, the model has reduced unnecessary spillage. Based on the model, five release strategies are prescribed dependent of the Lake Level forecast:

*Strategy 1, Saving strategy:*
If the Lake Level in the upcoming December is expected to be below the Lowest Regulated Water Level (LRWL of 473.5 m.a.s.l.), the release shall be according to a saving strategy. Release will be based on reduced hydropower production, which may be limited by the capacity curve (Figures 14 and 15).

*Strategy 2, Maximum production at hydropower plants:*
If Strategy 1 does not apply and Lake Level after 3 months is expected to be below the Highest Regulated Water Level (HRWL of 475.72 m.a.s.l.), this strategy can be followed. By applying this strategy, the release from Kamuzu Barrage shall be high enough to cover all the demand for all stakeholders, *i.e.* irrigation, water supply, and most predominant one, hydropower production.

*Strategy 3, Step 1 to avoid flooding:*
If Lake Level after 3 months is expected to increases to above the Highest Regulated Water Level (HRWL of 475.72 m.a.s.l.), but less than 475.85 m.a.s.l., a medium high flow shall be released, based on capacity curve or maximum 450 m$^3$/s. When the Lake Level is decreasing, the release will be reduced to maximum required if the Lake Level is expected to end below 475.72 m.a.s.l. after 3 months.

*Strategy 4, Step 2 to avoid flooding*
If the Lake Level after 3 months is expected to increase to above 475.85 m.a.s.l., but will be less than 476.1 m.a.s.l. (flood level) a medium high flow shall be released, based on capacity curve or maximum 900 m$^3$/s. When the Lake Level is decreasing the release will be reduced to 450 m$^3$/s if the Lake Level is expected to end below 475.85 m.a.s.l. after 3 months.

*Strategy 5, flood situation*
If Lake Level is expected during the next 3 months to increase to above el. 476.1 m.a.s.l., all gates are to be opened.

# Acknowledgement
Authors would like to acknowledge World Bank for providing the funding for the "detailed design of upgraded Kamuzu Barrage" projects. The contributions and feedbacks from Malawi's Ministry of Agriculture, Irrigation and Water Development, as well as the hydropower generation engineers of Electricity Supply Corporation of Malawi (ESCOM) are greatly acknowledged.